\documentclass[twocolumn,prl,showpacs,floatfix]{revtex4}

\usepackage{amsmath}
\usepackage{amsfonts}
\usepackage{amssymb}
\usepackage{bm}
\usepackage{color}
\usepackage{graphicx}

\begin{document}
\title{Stop-and-go kinetics in amyloid fibrillation}
  \author{Jesper Fonslet$^{1,2}$, Christian Beyschau Andersen$^{3,4}$, Sandeep Krishna$^1$, Simone Pigolotti$^1$, Hisashi Yagi$^{5,6}$, Yuji Goto$^{5,6}$, Daniel Otzen$^{7}$, Mogens H. Jensen$^{1\dag}$ and Jesper Ferkinghoff-Borg$^{8\dag}$}
\affiliation{$^1$Niels Bohr Institute, Blegdamsvej 17, DK-2100,
Copenhagen, Denmark}\homepage{http://cmol.nbi.dk}
\affiliation{$^2$Herlev Hospital, Klinisk Fysiologisk Afd. Herlev Ringvej 75, DK-2730.}
\affiliation{$^3$Novo Nordisk A/S, Protein Structure and Biophysics, Novo Nordisk Park, DK-2750 M\aa l\o v, Denmark}
\affiliation{$^4$National Research Council, Institute of Biophysics, Via Ugo La Malfa, 153, I-90146 Palermo, Italy}
\affiliation{$^5$Osaka University, Institute for Protein Research, Yamadaoka 3-2, Suita, Osaka 565-0871, Japan}
\affiliation{$^6$CREST, Japan Science and Technology Agency, Saitama, Japan}
\affiliation{$^7${\AA}rhus University, Department of Molecular Biology,
 Gustav Wieds Vej 10 C, 8000 {\AA}rhus C, Denmark}
\affiliation{$^8$DTU$\cdot$Elektro, Build. 349, \O rsteds Plads, Technical University of Denmark, 2800 Lyngby, Denmark}
\email{mhjensen@nbi.dk,jfb@elektro.dtu.dk}
\date{\today}

\begin{abstract}
  Many human diseases are associated with protein aggregation and
  fibrillation. We present experiments on in vitro glucagon
  fibrillation using total internal reflection fluorescence microscopy, providing real-time
  measurements of single-fibril growth. We find that amyloid fibrils grow
  in an intermittent fashion, with periods of growth followed by long
  pauses. The observed exponential distributions of stop and
  growth times support a Markovian model, in which fibrils shift
  between the two states with specific rates. Remarkably, the
  probability of being in the growing (stopping) state is very close to
  $1/4$ ($3/4$) in all experiments, even if the rates vary
  considerably. This finding suggests the presence of $4$ independent
  conformations of the fibril tip; we discuss this possibility in
  terms of the existing structural knowledge.
\end{abstract}
\maketitle

Protein fibrillation is the process by which misfolded proteins tend
to form large linear aggregates \cite{dobson}.
Its importance is related to the role
played in many degenerative diseases, such as Parkinson's, Alzheimer's,
Huntington and prion diseases \cite{chiti}. While our knowledge of the
structural properties of these fibrils improves at great pace
\cite{eisenberg,knowles2006,knowles2007}, the dynamics of their growth process is still poorly
understood.  The formation of amyloid fibrils involves at least two
steps: the formation of growth centers by primary nucleation, which is
often a slow process, followed by elongation through addition of
monomers \cite{huntington}. In many cases, a so-called secondary
nucleation mechanism is also involved, whereby new growth centers are
formed from existing fibrils
\cite{ferrone,librizzi,padrick,fodera,christian2}. Whereas the process
of secondary nucleation is known to entail a number of different mechanisms \cite{christian2},
the primary elongation process has not been elucidated to the same level of detail.

\begin{figure}[htb]
\includegraphics[width=8cm]{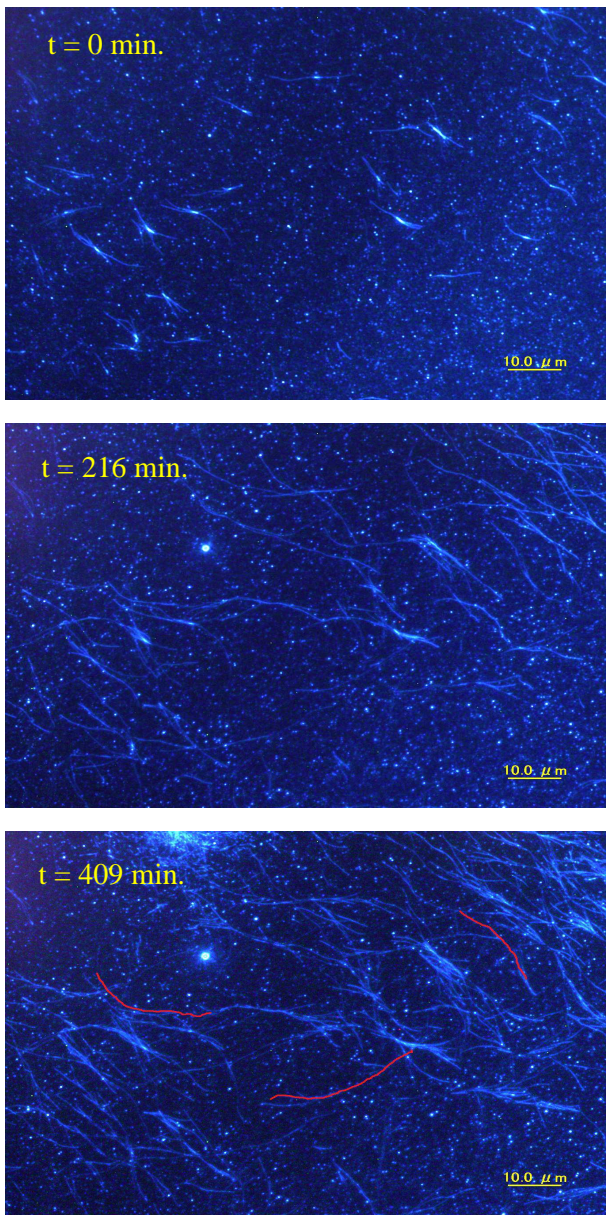}
\caption{TIRFM images of glucacon fibril growth with initial glucagon
concentration of $\rho=0.25~\mbox{mg/ml}$ in aqueous buffer (50 mM glycine HCl pH 2.5) at three consecutive times after the initiation of the
aggregation.  Red lines in the last picture mark examples of
  fibrils which are tracked during the growth process.}
  \label{tirfm}
\end{figure}

In this letter, we present an experimental and theoretical study of the elongation process of
glucacon fibrils. Glucagon is a small peptide hormone consisting of only 29 amino
acids produced in the pancreas. It has the opposite effect to
that of insulin and therefore increases blood glucose levels when
released. As a model system for protein fibrillation, glucagon
kinetics has provided insights into the early oligomerization stages
of the process \cite{svane,otzen-febs,otzen2008,otzen2009}, the interplay
between growth and fibril morphology \cite{pedersena,christian1} and amyloid
branching \cite{christian2}. Here, we focus on the properties of the
late-stage elongation process.

Experiments were performed on samples of glucagon monomers in
solution.  In order to detect the growth a specialized fluorescence
microscopy technique was applied, the so-called Total Internal
Reflection Fluorescence Microscopy (TIRFM). This technique utilizes total internal reflection to create
an evanescent electromagnetic field adjacent to the glass slide,
thereby exclusively exciting fluorophores in only a very thin
volume. The penetration depth, $d$, depends in a specific manner on
the wavelength and the angle of the incident light, as well as the
refractive indices of the media \cite{wazawa}. In the setup, an argon
laser was used along with a fused silica slide in contact with water,
leading to a penetration depth of $d=150~\mbox{nm}$. The TIRFM images
of the fibrillation process were obtained at initial glucagon
concentration of $\rho=0.25~~\mbox{mg/ml}$ in aqueous buffer (50 mM glycine HCl pH 2.5) with
preformed seeds. Images of the growth are shown in Fig. \ref{tirfm} at
three consecutive times $t=0,~216~,~407~~\mbox{min.}$ (see \cite{christian2} for
details on the experiment).

\begin{figure}[htb]
\includegraphics[width=9cm]{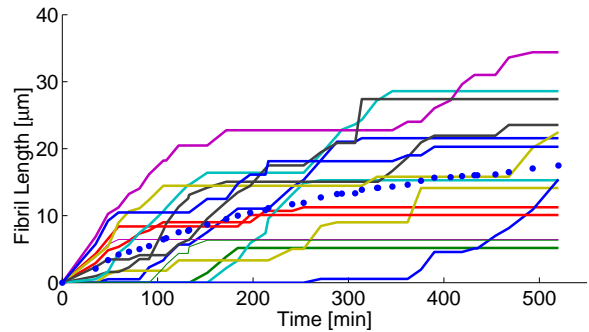}
\caption{Length as a function of time for $16$ fibrils tracked from the images shown in Fig 1. Note the long plateaus, corresponding
  to the stop states, followed by shorter (on average) growing
  periods. The average growth is indicated with a dotted blue
  line. The sampling time between each image can be seen from the time
  separation between each point.}
  \label{fib-data}
\end{figure}

Because the fibrils grow along the glass slide we are able to
track each fibril length as function of time. We monitor 16
independent fibrils for each image frame in the experiment. The time
interval, $\Delta t$, between frames varies from a minimum value
of $\Delta t_{min}=1~~\mbox{min.}$ to a maximum value $\Delta
t_{max}=35~~\mbox{min.}$, with a typical value of $\Delta
t=10~~\mbox{min.}$. The total duration of the experiment is
$t=525~~\mbox{min.}$ The combined results for the 16 fibrils are shown
in Fig. \ref{fib-data}.

\begin{figure}[htb]
\includegraphics[width=9cm]{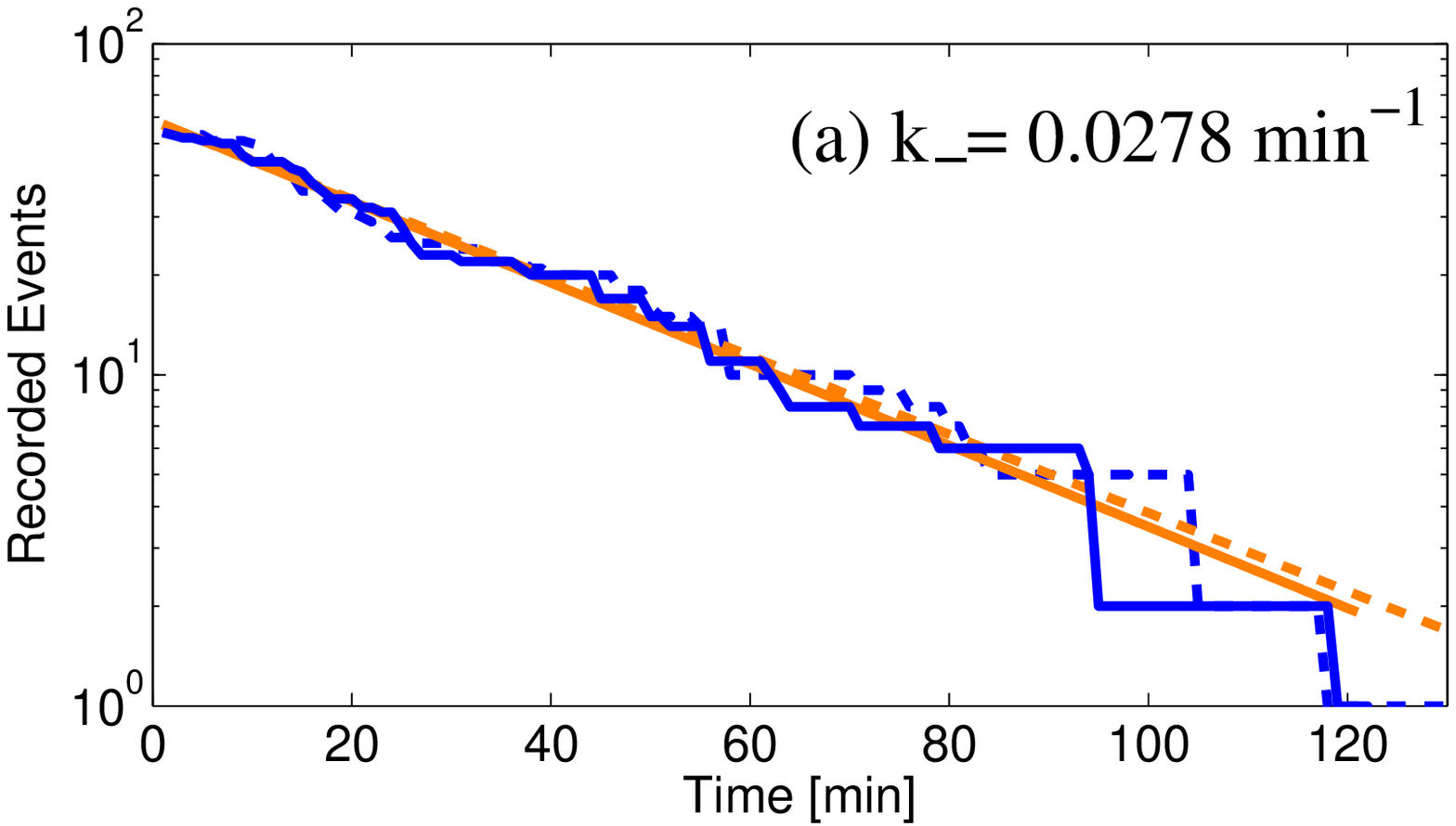}
\includegraphics[width=9cm]{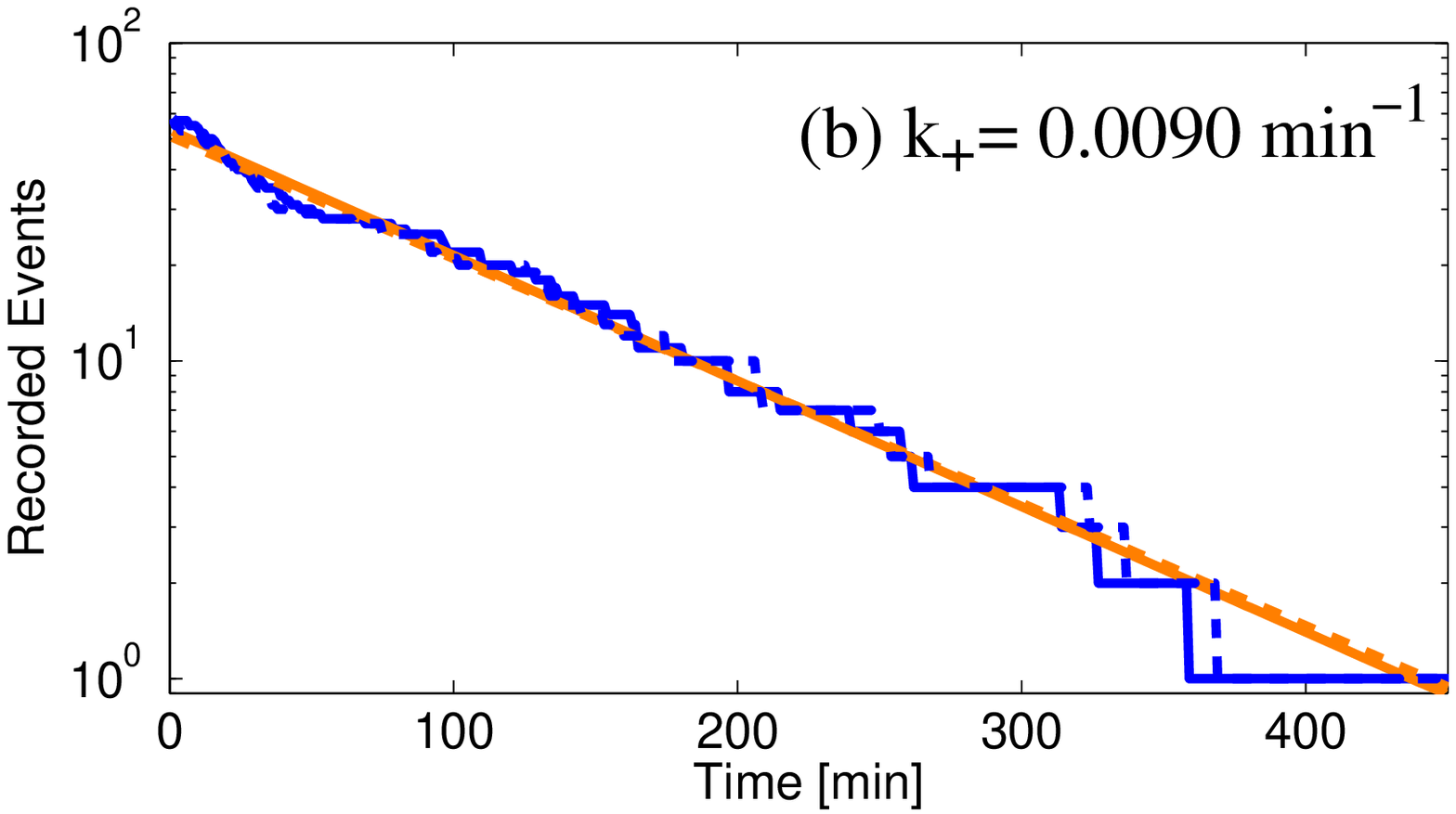}
\caption{Growth $f_g(t)$ (a) and stop $f_s(t)$ (b) times distributions
  on semi-logarithmic scales. Blue lines are data and yellow straight lines
  are the exponential fit. Both for data and fits, continuous lines
  are the lower estimates and dashed lines are the upper estimates
  (see text). All fits are of extremely good quality as indicated by the large R-value,
  $R^2>0.98$.}
  \label{exponential}
\end{figure}

A striking feature of the fibril dynamics is its discrete nature,
where long periods of growth are interrupted by extended periods of
stasis (stop state).  This prompted us to collect
the statistics of time spent in the growth (g) and the
stop (s) state, $f_g(t)$ and $f_s(t)$ respectively, for all 16
fibrils. In Fig. \ref{exponential}, these distributions are shown on semi-logarithmic plots.
Note that the finite sampling rate implies that actual time spent in given state can only be
estimated in terms of upper and lower bounds \cite{mogens}. The upper estimates are shown
by the dashed blue curves and the lower estimates are shown by the full blue curves. As seen in the figure,
the difference between the distributions for the upper and lower estimates of both $f_g$ and $f_s$ are marginal.
All distributions are very well fitted by exponential functions, $f_g(t)\sim \exp(-k_- t)$ and $f_s(t)\sim \exp(-k_+ t)$,
as shown by the yellow curves (here, the dashed yellow curve is the fit to the upper estimates and the full yellow curve is the fit of  the lower estimates). The fits are of excellent quality over almost two decades, as signified by high $R$ values ($R^2>0.98$).  The values obtained are $k_+=9.0\cdot 10^{-3}~~\mbox{min}^{-1}$ and $k_-=2.8\cdot 10^{-2}~~\mbox{min}^{-1}$.
A series of four independent experiments $A-D$ have been performed with the same glucagon monomer
concentration (0.25 mg/ml) and pH (2.5) with differences in the seed concentrations ($\sim$20\% variation) and data acquisition only. Specifically, the maximal frame length was varied by a factor 3 and the minimum frame length by a factor 18.
In all four cases the same analysis and procedure was performed resulting in exponential
distributions of similar high quality. The results are summarized in
Table \ref{table_data}.

\begin{figure}[htb]
\includegraphics[width=9cm]{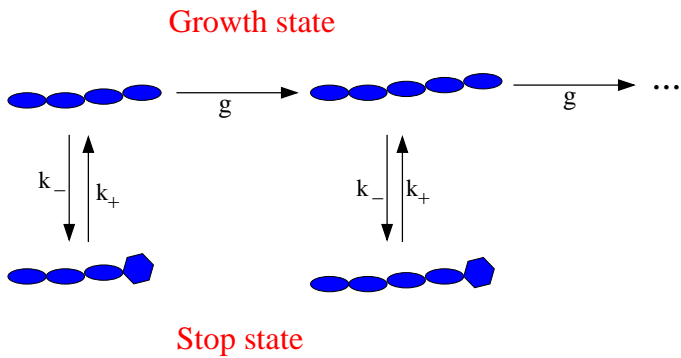}
\caption{The simplest two-state model of fibril growth. The fibril is
  assumed to be in one of two states with intrinsic transition rates
  of $k_+$ and $k_-$. In the growing state, monomer attachment occurs
  with the rate $g$.}\label{sandeep}
\end{figure}

The observed stop and go behaviour of fibril dynamics is clearly
not associated with the discrete nature of monomer attachment \cite{diffusion}.
The simplest model of the process is to assume that
the fibril exhibits two internal states: one in which it is allowed to
grow, with a rate $g$ and one in which it cannot grow. The intrinsic
transition rates between the two states is then identified with the
observed transition rates $k_+$ (stop$\rightarrow$ growth) and $k_-$
(growth $\rightarrow$ stop), see Fig. \ref{sandeep}.

The Markovian nature of the model implies, that the time spent in each
state is exponentially distributed which is indeed consistent
with the results from
the data-analysis, Fig.(\ref{exponential}), leading to estimates
of $k_+$ and $k_-$. Denoting the total rate $k=k_++k_{-}$, one can
calculate the probability of being in the growing state, $p_+=k_+/k$
and in the stopped state, $p_-=k_-/k$. Table \ref{table_data} presents
a summary of the parameters in $4$ different experiments. By comparing the
four different series, we observe that the rates $k_+$ and $k_-$
vary quite significantly, up to a factor 3.  A possible explanation for this
could be the variation in sampling frequency (as is indicated in Table \ref{table_data}),
due to heating of the sample by the laser. Indeed, the dependency of the fibril growth
on both the laser intensity as well as the illumination time has been observed under similar experimental conditions for $\beta_2$-microglobulin kinetics \cite{ozawa}.

The striking result is that, even though both transition rates vary among experiments, they combine in such a way that the probabilities
of growing and stopping, $p_+$ and $p_-$, do not change appreciably in
different experiments. In particular, $p_+$ is always very close to
$1/4$ and consequently $p_-$ is very close to $3/4$. Notice that, if
the difference between the growing and the stopped state would have
been due to an energy gap, one would have expected the population
ratio of the two states to be much more sensitive to variations
in the individual transition rates, $k_\pm$. Conversely, the constancy
of this ratio suggests that the energy difference between the growing and
the stopped states to be irrelevant. This ratio could then reflect the presence of
three stopped configurations for each growing one, all of them being isoenergetic.

One may wonder whether the observed '$1/4$-$3/4$' law is really independent of the
monomer concentration, as in the model. Due to experimental limitations, this has not been directly tested. However,
the observed average growth rate, $g$, displays more than a two-fold variation from Exp. A to Exp. D,
suggesting notable differences in the local monomer concentration of the growing fibrils.
If the state probabilities were sensitive to changes in the concentration, we would have expected a correlation
between $p_\pm$ and $g$. We further tried to test the pertinence of the state probabilities by dividing the data sets into two parts, one corresponding to the first half of the experiment and the other to the second half, and measuring
$p_+$ and $p_-$ separately in the two parts. Unfortunately, the
statistical noise increased significantly when considering sub-parts
of the data series and neither added to nor altered our conclusions.

To conclude, we have presented a stop-go model for glucagon fibril elongation. In fact, similar kinetic behavior has recently been reported for the fibril elongation of $A\beta$ -peptides \cite{kellermayer, ban2004} as well as for $\alpha$-synuclein \cite{hoyer},
although the timescales involved are 1-2 orders of magnitude faster. This suggests that the observed stop-go kinetics reflects
the presence of some kind of structural change at the fibril ends, which is not necessarily specific to glucagon. In this picture, the protein properties would affect the barrier height, and thus the timescale of the process, only.

If our hypothesis about the existence of approximately
isoenergetic states holds up to scrutiny, it suggests that
there may be an additional dimension in the fibrillation
energy landscape that cannot easily be identified by
conventional techniques. Glucagon is known to adopt a number
of different  conformations depending on the fibrillation
conditions, but these conformations differ considerably in
energy and are unlikely to co-exist to an equal extent \cite{pedersen-otzen}
. Rather, it is possible that we have a number of closely
related states with different propagation properties which
are separated by high local activation barriers within a
relatively flat ground state level. Further experimental
studies are required to establish the validity of this
suggestion.

This research has been supported by the VILLUM KANN RASMUSSEN
Foundation and the Danish National Research Foundations. We are
grateful to Christian Rischel and Joachim Mathiesen
for discussions at the early stage of this work.

\begin{table}
\begin{center}
\begin{tabular}{|l|l|l|l|l|}
\hline
{\bf Parameters} & Exp. A & Exp. B & Exp. C & Exp. D \\
\hline
\hline
$k_+\ [\mathrm{min}^{-1} ]$  & 0.0090 & 0.0192 & 0.0075  & 0.0059  \\
& $\pm 0.0001$ & $\pm$ 0.0004 & $\pm$ 0.0001 & $\pm$ 0.0001 \\
\hline
$k_-\ [\mathrm{min}^{-1} ]$  & 0.0278 & 0.04780  & 0.023  & 0.0166  \\
& $\pm$ 0.0005 & $\pm$ 0.00002 & $\pm$ 0.002 & $\pm$ 0.0002 \\
\hline
$p_+$                       & 0.244 & 0.249 & 0.247 & 0.260  \\
\hline
$p_-$                       & 0.756 & 0.751 & 0.753 & 0.740  \\
\hline
g [nm/min] & 135 & 116 & 111 & 56 \\
\hline
\hline
Tot. time [min] & 525 & 340 & 630 & 1030  \\
\hline
Max. FL  [min]  & 35 & 27 & 24 & 90  \\
\hline
Min. FL  [min]  & 1 & 3 & 1 & 18  \\
\hline
Avr. FL  [min]  & 13.7 & 11.8 & 20.7 & 30.9  \\
\hline
\end{tabular}
\end{center}
\caption{Estimated kinetic parameters (row 1-5) and parameters for the data acquisition (row 6-9) in
$4$ different experiments, A-D. FL stands for frame length. All figures of this paper refer to the
experiment of the first column, Exp. A. The average growth rate of the fibrils over the total time of each experiment is indicated by
$g$ (no appreciable effect of depletion is observed). These values are comparable to the measurements in \cite{christian2}.}
\label{table_data}
\end{table}


\begin{thebibliography}{50}
\bibitem{mogens} The finite sampling frequency implies that the exact
  time of a state transition (growth $\leftrightarrow$ stop) can be
  anywhere in the interval between two observations. Therefore, an
  upper and lower bound of the time spent in a particular state is simple given by
  choosing either the 'outer' or the 'inner' observation points.
\bibitem{diffusion} The monomer attachment rate, $r$, is approximately given by
$r\simeq g\pi R^2/V_{glu}\sim 10^3~\mbox{min.$^{-1}$}$, where $g\sim 100~\mbox{nm/min.}$ is the average observed elongation rate (see table \ref{table_data}), $R\sim 5~\mbox{nm}$ is the fibril radius and $V_{glu}\sim 4~\mbox{nm$^3$}$ is the volume of a fibrillated glucagon peptide. This rate is several orders of magnitude larger than the observed transition rates.
\end{thebibliography}
\end{document}